\documentclass[aps,physrev,preprint,groupedaddress]{revtex4-2}

\usepackage{graphicx}
\usepackage{aas_macros}

\begin{document}


\title{Influence of effective mass of the relativistic mean field theory \\
on core collapse supernovae and compact objects}


\author{Kohsuke Sumiyoshi}
\email{sumi@numazu-ct.ac.jp}
\affiliation{National Institute of Technology, Numazu College,
3600 Ooka, Numazu, Shizuoka 410-8501, Japan}

\author{Hajime Togashi}
\affiliation{Department of Physics, Kyoto University, 
Kita-shirakawa Oiwake-cho, Sakyo-ku, Kyoto 606-8502, Japan}

\author{Shun Furusawa}
\affiliation{College of Science and Engineering, Kanto Gakuin University, 
1-50-1 Mutsuura-higashi, Kanazawa-ku, Yokohama, Kanagawa 236-8501, Japan}

\author{Shuying Li, Hong Shen}
\affiliation{%
 School of Physics, Nankai University, 
 Tianjin 300071, People's Republic of China
}%

\author{Ken'ichiro Nakazato}
\affiliation{Faculty of Arts and Science, Kyushu University, 
744 Motooka, Nishi-ku, Fukuoka 819-0395, Japan}

\author{Hideyuki Suzuki}
\affiliation{Faculty of Science \& Technology, Tokyo University of Science, 
2641 Yamazaki, Noda, Chiba 278-8510, Japan}


\date{\today}

\begin{abstract}
We study the influence of the effective mass in the relativistic mean field (RMF) theory on the properties of the central core of collapse-driven supernovae and the formation of compact objects.  
Influence of the effective mass has been so far studied within the non-relativistic frameworks.  
In order to clarify the role of the effective mass in the relativistic frameworks, which is different from non-relativistic ones, 
we adopt the set of equation of state (EOS) tables using the parameterizations TM1e and TM1m, which have different effective masses but with the same saturation properties, in the RMF theory.  
We show that choices of the effective mass in supernova matter affect both the stiffness of the EOS through pressure and the thermodynamical behavior through temperature under the RMF frameworks.  
We explore differences in matter evolution with neutrino emissions by performing a set of numerical simulations of the gravitational collapse and bounce of massive stars and the cooling of the proto-neutron stars.  
The EOS with large effective mass leads to compact proto-neutron stars and early collapse to black holes with high densities and temperatures due to the softness.  
It leads to high energy neutrinos in long emission from the proto-neutron star cooling and in short burst from the black hole formation.  

\end{abstract}


\maketitle

\section{Introduction
\label{Intro}}

Properties of hot and dense matter play an essential role in understanding the phenomenon of core-collapse supernovae as the end of the life of massive stars \cite{oer17,sum23book}.  
They determine the dynamics of gravitational collapse of the central Fe core, launch of the shock wave by the core bounce, and subsequent propagation, which may or may not lead to a successful explosion \cite{bet90}.  
In a modern explosion scenario, neutrino heating is an important agent to assist the revival of stalled shock waves \cite{bet85} together with hydrodynamical instabilities \cite{jan12a,bur13,kot13,jan16,jan17b,jan25}.  
The equation of state (EOS) may affect the explosion in multi-dimensional simulations \cite{jan23book} in various ways, but its influence on the explosion dynamics has not yet been fully clarified.  
As a general tendency, the softness of EOS leads to a favorable condition to have more neutrino heating for the explosion, since neutrino emission is enhanced in the compression of matter at high density and temperature inside the compact object.  
Although there has been progress to assess the EOS of cold neutron stars through observational and experimental constraints \cite{lat21,burg21,chat25}, the properties of dense matter at finite temperature in the supernova core remain uncertain, and therefore assessment of the impact of hot and dense matter on the supernova mechanism is important for modern numerical studies of supernovae.  

The EOS is also crucial in determining the properties of the birth of compact objects, a proto-neutron star or black hole, depending on the fate of explosion dynamics.  
The outcome of nascent compact object is determined by the maximum mass supported by the EOS of hot and dense matter, which is different from the one of cold neutron stars. 
The formation of compact objects is associated with the emission of neutrinos and gravitational waves, which can be used to probe the hot and dense matter inside the compact objects \cite{kot06,mir16,mul19,abd20,mul26,so26}.  
The softness of the EOS, for example, may lead to enhanced fluxes of neutrinos with a high average energy due to strong compression of matter at high density and temperature.  
To explore the influence of hot and dense matter on supernova dynamics and nascent compact objects, there have been many efforts to provide the supernova EOS for numerical simulations by utilizing analytic formulae or constructing the data table of the quantities of the EOS \cite{oer17,sum23book}.  
There has been extensive research to compare the outcome of supernova dynamics with different sets of the data tables of supernova EOS and to explore systematic trends by changing the parameters of the EOS in the analytic formulae in spherical \cite{bar85,tak88,swe94,tho03,sum05,fis11,ste13} and multi-dimensional \cite{mar09b,suw13,cou13,fis14,sher17b,bur18,nag18,har19,sch19b,ande21,boc22,pow25,rus26} simulations.  
There has also been research on the dependence of black hole formation on the EOS variations \cite{sum06,sum07,ocon11,char15,pan18,sch20,ande25}.  



We focus on influence of the effective mass of nucleons in hot and dense matter in this study.  
The effective mass of nucleons is influential in determining the thermodynamical properties of dense matter at finite temperature through the kinetic term of single particle energies and the behavior of nuclear interaction.  
In the previous study using the non-relativistic framework based on the Skyrme interaction \cite{yas18}, 
it is claimed that large effective masses are preferable in the explosion dynamics through reduction of pressure and a rapid contraction of the proto-neutron star.  
More compact proto-neutron stars lead to higher density and temperature of matter, which enhances the emission of neutrinos from the surface of proto-neutron star.
In the neutrino heating mechanism for the explosion, 
increase in the heating rate via neutrino absorption on nucleons behind the shock wave may assist to turn the stalled shock wave into outwardly propagating one for the explosion.  
It has been shown that the revival of the stalled shock wave occurs earlier with the EOS with larger effective masses in more systematic numerical simulations \cite{sch19b,ande21}.  
They attribute the preference of large effective mass to the decrease of the pressure of nucleons through the non-relativistic kinetic term.  

We note that these previous studies are limited within the EOS models by non-relativistic frameworks such as the one using the Skyrme-type interaction \cite{sch19a,tog25}.  
The influence of the effective mass using the EOS models in relativistic framework has not been studied well so far.  
In the non-relativistic models, the description of the effective mass of nucleons relies solely on the functional form of kinetic term and is independent of other forms of potential terms.  
In the relativistic mean field theory, on the other hand, the effective mass is obtained in a self-consistent manner from the effective lagrangian so that both kinetic and potential terms are influential in the behavior of EOS.  
It is hence indispensable to examine the influence of effective mass using the EOS evaluated in the relativistic mean field theory and to explore its impact on core-collapse supernovae.  

%


The effective mass is also influential in determining the properties of supernova neutrinos, which are important for the prediction of neutrino bursts at observational facilities.  
Numerical simulations of the thermal evolution of proto-neutron stars born after the core bounce have been performed to explore the properties of hot and dense matter \cite{bur86,suz94,sum95c,pon99,rob12,nak13a}.  
The neutrinos are once trapped inside the central core and later diffuse out from the surface of proto-neutron star.  
The properties of neutrino emission reflect the thermodynamical conditions of the proto-neutron stars, which are determined by the EOS.  
In the systematic numerical studies, the effective mass is influential to determine the cooling timescale of the proto-neutron star and its associated properties of neutrino emission \cite{nak19}.   
It has been shown that large effective masses lead to long duration of neutrino bursts due to slow emission of neutrinos from proto-neutron star with high density and temperature.  
In their study, the analytic formulae for the thermal contributions have been used in the expression of the non-relativistic kinetic term with the effective mass of nucleons.  
The studies are again limited within the non-relativistic frameworks and studies with the relativistic frameworks are necessary.  

We explore the influence of the effective mass in the relativistic mean field (RMF) theory in this study.  
The effective mass of nucleons in the relativistic many body framework is defined in the terms of the mass term in the Dirac equation.  
The effective mass in the RMF theory is described via attractive interaction of the scalar meson and determined by the form of the effective Lagrangian.  
The role of the effective mass is different from the non-relativistic frameworks such as the Skyrme-interaction, which is used to derive some sets of the supernova EOS.  
Therefore, we examine the hot and dense matter to reveal the impact of different effective masses in the RMF theory on supernovae and compact objects.  


Recently, the new EOS table (TM1m) has been constructed \cite{sli25} to assess the influence of the effective mass based on the EOS table (TM1e) \cite{she20}, which is the modern standard set in the series of the Shen EOS tables \cite{she11}.  
The TM1m interaction is determined to set a large effective mass compared to TM1e by keeping other saturation properties so that we can solely examine the influence of the effective mass \cite{sli25}.  
We adopt the two EOS tables, TM1m and TM1e, in a set of numerical simulations to explore the difference in the core-collapse and bounce, the cooling of proto-neutron stars, and the formation of black hole.  

In this article, we examine the properties of supernova matter with leptons and radiation 
to examine the basic difference in the thermodynamical quantities.  
In order to assess the impact of the effective mass on the phenomena of core-collapse supernovae, 
we perform numerical simulations of the general relativistic neutrino-radiation hydrodynamics under the spherical symmetry to examine the size of differences in the collapse and bounce of the supernova core and to explore the impact in the black hole formation.  
We also perform numerical simulations of the cooling of the proto-neutron star to examine the difference in the signals of the supernova neutrinos.  

This study of the influence of EOS on supernovae is a work along the lines of systematic studies using the Shen EOS tables.  
The influence of the symmetry energy using the EOS with interaction TM1e and TM1 \cite{she98a,she98b,she20} has been studied in Ref. \cite{sum19}.  
We follow the basic procedure in the previous numerical studies to perform new simulations so that one can easily compare with the former models.  

This paper is arranged as follows.
We describe the basic properties of dense matter in the RMF theory with different effective masses in \S \ref{sec:RMF}.  
Starting with the properties of matter at zero temperature in \S \ref{sec:Matter}, we describe the behavior of supernova matter at finite temperature \S \ref{sec:SNmatter}.  
We briefly describe the setting of numerical simulations in \S \ref{sec:massive_star}.
We report numerical results of core collapse and bounce in \S \ref{sec:post_bounce} and black hole formation in \S \ref{sec:BH_formation}.  
We describe the numerical setting and report numerical results of proto-neutron star cooling in \S \ref{sec:PNS_cooling}.  
We summarize the article with discussion in \S \ref{sec:Summary}.  


\section{Sets of equation of state in the RMF theory
\label{sec:RMF}}

We explore the influence of the effective mass 
by using the two sets of the EOS derived in the RMF theory.  
We adopt the EOS table with the TM1e interaction \cite{she20} as the basis for comparison and utilize the EOS table newly constructed with the TM1m interaction with a different effective mass \cite{sli25}.  
Although the original version of the EOS table with the TM1 interaction \cite{sug94,she98a,she98b,she11} has been widely used so far in numerical simulations of supernovae, the EOS table with the TM1e interaction has been constructed by updating the size of the symmetry energy to satisfy the experimental and observational constraints and to explore the effect of the symmetry energy.  
The EOS table with the TM1m interaction has recently been constructed by choosing a large effective mass for comparison based on the TM1e interaction while maintaining the other saturation properties.  
Detailed descriptions about the two EOS sets with TM1e and TM1m can be found in \cite{sli25}.  

In this study, we focus on the differences in the phase of uniform matter at high densities.  
We commonly adopt the EOS of TM1e at sub-saturation densities for the description of non-uniform matter.  
We connect the TM1m EOS for uniform matter with the TM1e EOS for non-uniform matter when we need to cover the wide range of densities.  

\subsection{Nuclear matter and neutron stars
\label{sec:Matter}}

\begin{figure}
\includegraphics[width=8.5cm]{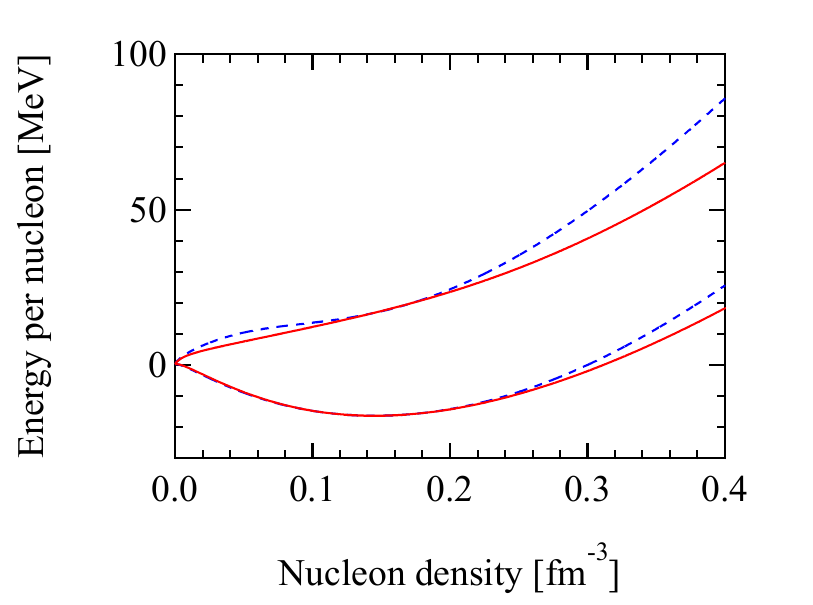}%
\includegraphics[width=8.5cm]{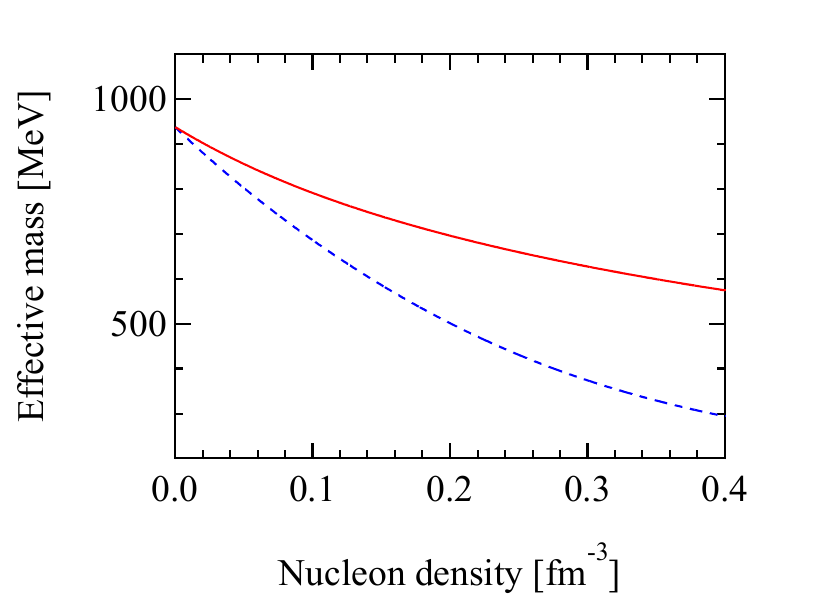}%
\caption{Properties of uniform matter of nucleons in the cases of TM1m (red, solid) and TM1e (blue, dashed) are plotted as functions of nucleon density.  
The energies per nucleon of symmetric nuclear matter ($Y_p=0.5$) and neutron matter ($Y_p=0.0$) and the effective mass of symmetric nuclear matter ($Y_p=0.5$) are shown in the left and right panels, respectively.  
\label{fig:Nmatter_EOS}}
\end{figure}

We briefly describe the basic properties of the nuclear matter and neutron star matter for TM1e and TM1m EOSs at zero temperature for reference in further astrophysical applications.  
In Fig. \ref{fig:Nmatter_EOS}, we show the behavior of energies and effective mass of uniform nuclear matter as a function of nucleon number density.  
Properties at the saturation densities are listed in Table. \ref{tab:EOS}.  

The energy per nucleon of symmetric nuclear matter with the proton fraction $Y_p=0.5$ for TM1m is nearly the same as the corresponding one for TM1e around the common saturation density of 0.145~fm$^{-3}$ and energy of $-16.3$ MeV.  
The value of the incompressibility, $K$, for TM1m is also the same as that for TM1e.  
The effective mass, $M^{*}$, for TM1m of symmetric nuclear matter is larger by choosing the weaker attraction via the scalar meson coupling with nucleons than that for TM1e.  
The large effective mass leads to the reduction of kinetic energy of nucleons and is linked with the reduction of the repulsion via the vector meson coupling with nucleons (see discussion below).  
These differences result in a slow increase of the energy in symmetric nuclear matter at high densities and make TM1m EOS softer than TM1e EOS.  

The energy curves of pure neutron matter follow closely around the saturation density with the same value of the symmetry energy, $E_{sym}$, and the symmetry energy slope, $L$.  
The energy for TM1m is lower than that for TM1e, having the larger effective mass and the weaker repulsion in a similar manner to the case of symmetric nuclear matter.  
This behavior leads to the softness of the neutron star matter in TM1m EOS and its neutron star properties, as we will show below.  

\begin{table}[htp]
\caption{Properties of nuclear matter at the saturation density for TM1m and TM1e EOSs.  The values of the incompressibility, $K$, the symmetry energy, $E_{sym}$, and its slope, $L$, and the effective mass ratio, $M^{*}/M$, are listed.  }
\begin{center}
\begin{tabular}{ccccc}
\hline\hline
EOS &\; $K$[MeV] &\; $E_{sym}$[MeV] &\; $L$[MeV] &\; $M^{*}/M$ \\
\hline 
TM1m & 281 & 31.4 & 40 & 0.793 \\
TM1e & 281 & 31.4 & 40 & 0.634 \\
\hline
\end{tabular}
\end{center}
\label{tab:EOS}
\end{table}%

\begin{figure}
\includegraphics[width=8.5cm]{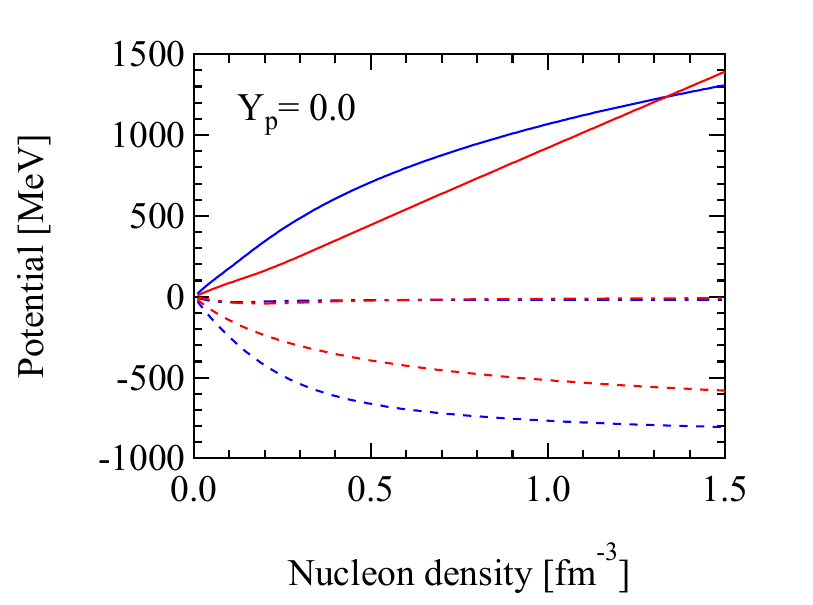}%
\includegraphics[width=8.5cm]{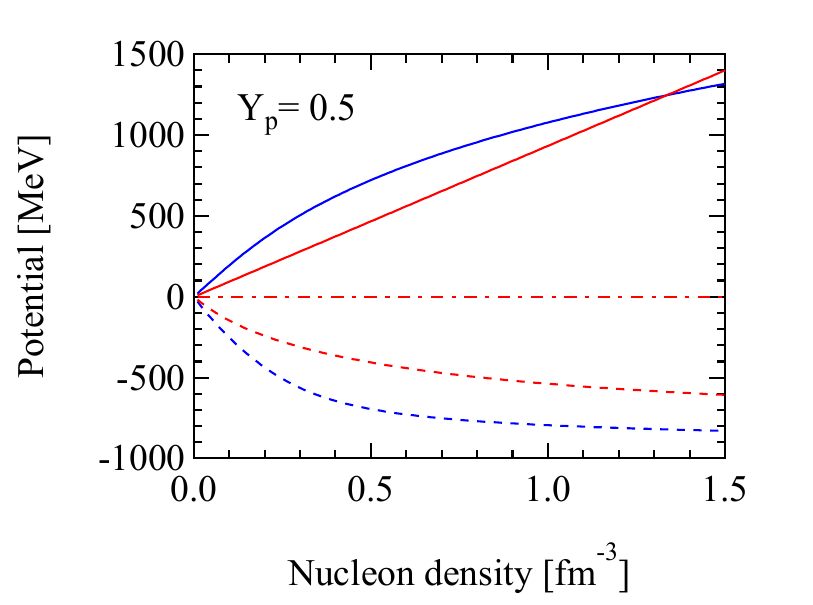}%
\caption{Mean field potentials of neutron matter ($Y_p=0.0$) and symmetric nuclear matter ($Y_p=0.5$) are shown as functions of nucleon density in the left and right panels, respectively.  
The isoscalar-vector, isovector-vector, and isoscalar-scalar potentials in the cases of TM1m (red) and TM1e (blue) are plotted by solid, dash-dotted, and dashed lines, respectively.  
\label{fig:Nmatter_EOS_Upot}}
\end{figure}

We note that the choice of a large effective mass in the RMF theory is closely linked with the behavior of the energy at high densities.  
This feature originates from the mechanism of saturation through the balance between strong attractive and repulsive interactions commonly seen in relativistic many body frameworks such as the RMF theory.  
We show in Fig. \ref{fig:Nmatter_EOS_Upot} the density dependence of vector and scalar potentials in the Dirac equation for nucleons evaluated for neutron matter and symmetric nuclear matter in the RMF theory.  
A strongly attractive potential brought by the scalar meson provides the effective mass smaller than the bare nucleon mass as the mean field in nuclear matter.  
The scalar potential for TM1m is shallower than that for TM1e.  
The corresponding vector potential is repulsive in providing the saturation properties and depends on the strength of attraction of the scalar potential.  
The repulsion of vector potential for TM1m is weaker than that for TM1e due to the difference in the scalar potential, i.e. the effective mass.  
It is interesting to see that the vector potential for TM1m grows linearly as the density increases, which is constrained by the reduced repulsion at the saturation density, and it overcomes the case of TM1e at high densities $\geq1.3$~fm$^{-3}$.  
The behavior of the energy through the potentials at high densities is determined by the terms of mean fields for the vector meson including non-linear terms in the adopted lagrangian of the RMF theory, but is constrained by the settings of the nuclear properties at the saturation density.  
This behavior of EOS in the RMF theory is apparently different from the counter part EOS of non-relativistic frameworks such as the one using the Skyrme interaction.  

\begin{figure}
\includegraphics[width=9cm]{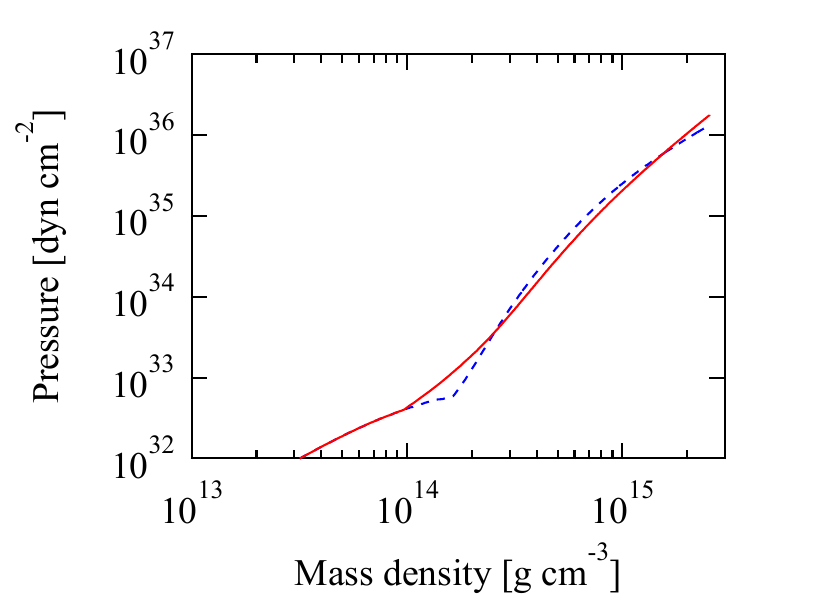}%
\caption{The pressure of cold neutron star matter in the cases of TM1m (red, solid) and TM1e (blue, dashed) is plotted as a function of mass density.  
\label{fig:NSmatter_NS_EOS}}
\end{figure}

\begin{figure}
\includegraphics[width=9cm]{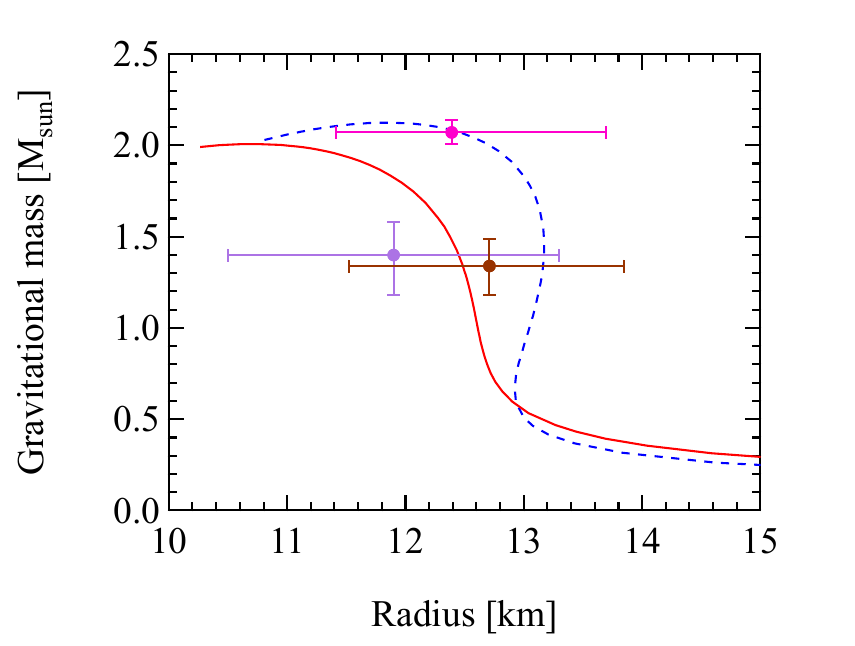}%
\caption{The mass-radius relations in the case of TM1m (red, solid) and TM1e (blue, dashed) are shown with the observational data by dots and lines for GW170817 \cite{abb18} (violet), PSR J0030+0451 \cite{ril19} (brown) and PSR J0740 +6620 \cite{ril21} (magenta).  
\label{fig:NSmatter_NS_MR}}
\end{figure}

The difference in nuclear matter between TM1m and TM1e is reflected in different neutron star profiles.  
In Fig. \ref{fig:NSmatter_NS_EOS}, we show the pressure of the cold neutron star matter calculated for the cases of TM1m and TM1e as a function of the mass density.  
The increase of pressure for TM1m is slow above the saturation density, $3\times10^{14}$~g\,cm$^{-3}$ due to its softness of neutron matter.  
It becomes rapid at high densities above $\sim10^{15}$~g\,cm$^{-3}$ due to the dominance of the repulsive vector potential as explained above.  
We show in Fig. \ref{fig:NSmatter_NS_MR} the gravitational mass of neutron stars as a function of radius for the cases of TM1m and TM1e.  
The radii of neutron stars for TM1m are smaller than those for TM1e due to its softness.  
The maximum masses for TM1m and TM1e are 2.01 and 2.12$M_{\odot}$, respectively.  
Note that we adopt the same EOS by TM1e for cold neutron star matter in crust regions at low densities to discuss solely the difference of uniform matter at high densities.  

\subsection{Supernova matter
\label{sec:SNmatter}}

We examine the thermal properties of supernova matter to understand differences in the two EOS sets at finite temperature before we discuss the results of core-collapse supernovae and proto-neutron stars.  
We evaluate properties of the supernova matter composed of nucleons, electrons, positrons, and photons by adding the lepton and photon contributions to the nuclear contributions for pressure, energy, and entropy.  
We set the condition of the supernova matter by fixing the entropy per baryon, $S=1$~[$k_B$/baryon], and the net electron fraction, $Y_e=0.3$, which are typical in supernova cores.  

\begin{figure}
\includegraphics[width=8.5cm]{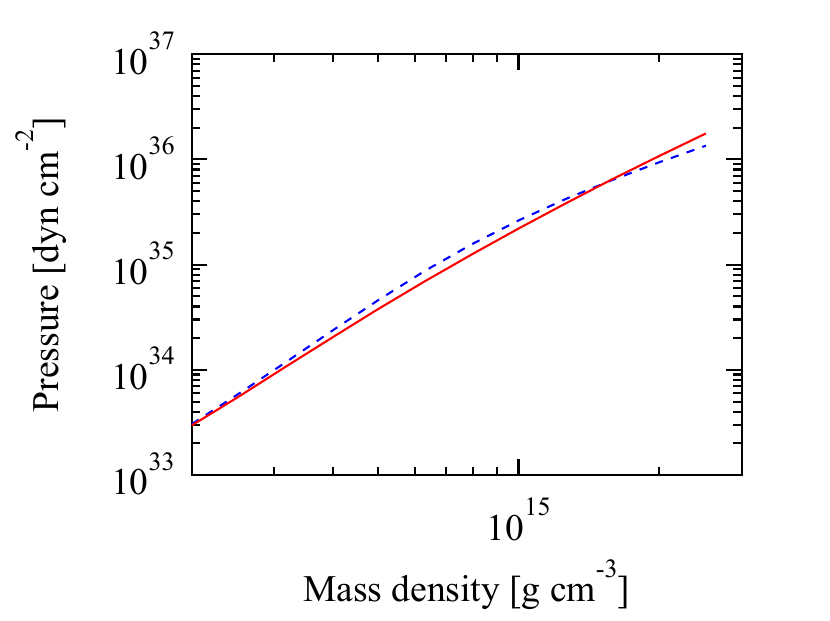}%
\includegraphics[width=8.5cm]{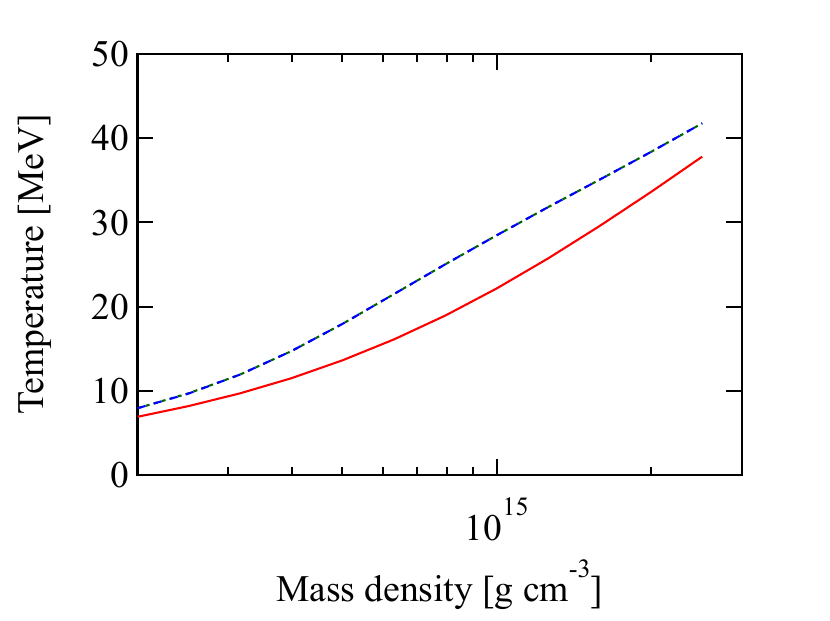}\\%
\includegraphics[width=8.5cm]{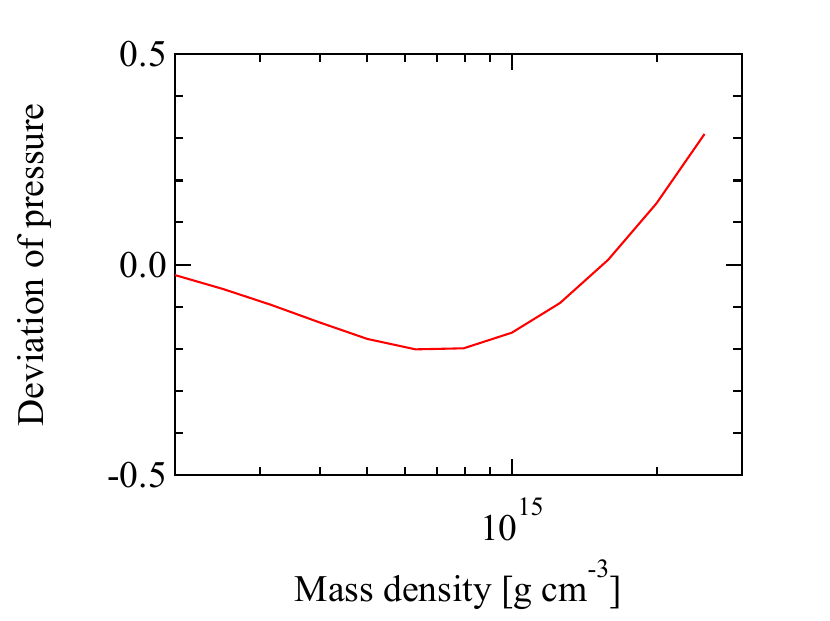}%
\includegraphics[width=8.5cm]{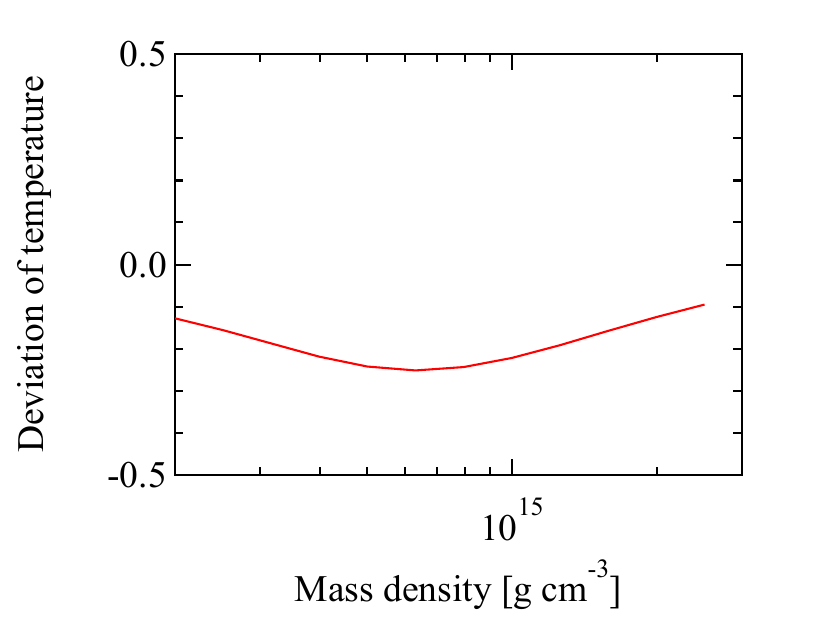}%
\caption{Thermodynamical quantities of the hot and dense matter in the case of TM1m (red, solid) and TM1e (blue, dashed) are shown as functions of the mass density for the fixed values of $S$=1~[$k_B$/baryon] and $Y_e=$0.3.  
Pressure and temperature are displayed in the upper-left and upper-right panels, respectively.  
Relative deviations of pressure and temperature of TM1m with respect to those of TM1e are displayed in the lower-left and lower-right panels, respectively.  
\label{fig:SNmatter_EOS}}
\end{figure}

We show in Fig. \ref{fig:SNmatter_EOS} the comparison of the thermal properties of the evaluated matter and their relative differences as functions of the mass density.  
The pressure for TM1m is lower than that for TM1e up to $\sim10^{15}$~g\,cm$^{-3}$ and is higher for high densities.  
The relative deviation of the pressure amounts to $\sim$ 20\% at maximum.  
This behavior of softness and stiffness is similar to the one seen in nuclear matter and neutron star matter.  
The temperature for the fixed entropy per baryon for TM1m is lower by $\sim$ 20\% than that for TM1e.  
This is because the entropy per baryon for TM1m is larger than that for TM1e at fixed temperature due to the large effective mass.  
Large effective masses lead to high level density of the energy spectra in the kinetic term of nucleons and results in high entropy.  
The low temperature in TM1m may affect the thermal properties of supernova cores and proto-neutron stars and change neutrino energies.  
This reduction may be influential for the neutrino heating mechanism and the thermal emission of supernova neutrinos.  
We note that there can be enhancement of temperature due to compression of matter, hence, it is mandatory to perform numerical simulations to find out net effects as the whole.  

\begin{figure}
\includegraphics[width=8cm]{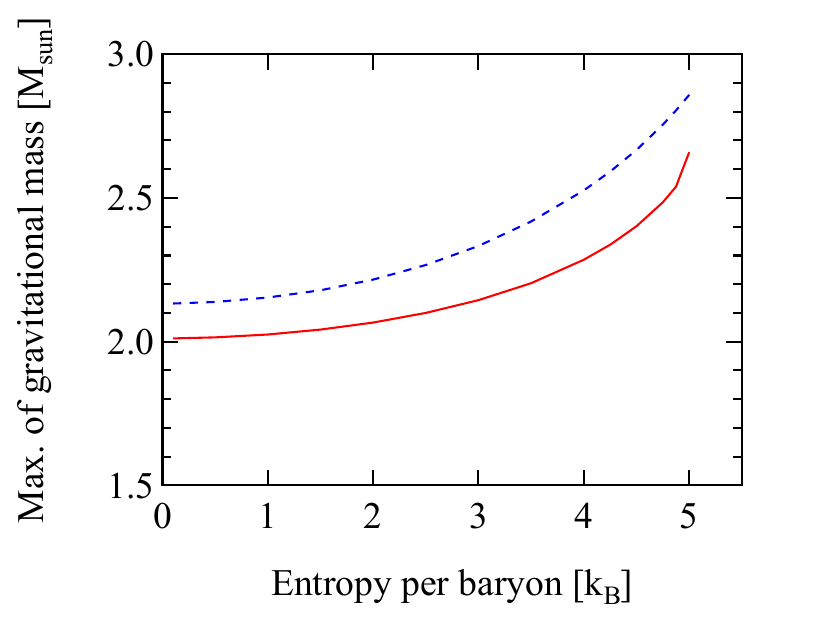}%
\caption{Maximum mass of the proto-neutron stars evaluated by adopting hot and dense matter composed of nucleons with electrons, positrons, and photons under the beta equilibrium for various fixed values of entropy per baryon (see text for detailed setting).  
The maximum of the gravitational masses are plotted as a function of the entropy per baryon in the cases of TM1m (red, solid) and TM1e (blue, dashed).  
\label{fig:protoNS_Mmax}}
\end{figure}

The difference in the stiffness of two EOS sets can be influential in the formation of black holes from massive proto-neutron stars \cite{sum06,sum07,ocon11,sch20}.  
When the mass of the massive proto-neutron stars with matter accretion reaches the critical mass supported by the hot and dense matter, they collapse to the black hole with termination of neutrino emission.  
We show in Fig. \ref{fig:protoNS_Mmax} the maximum mass of the proto-neutron stars determined from the sequence of the hydrostatic structure adopting the EOS for supernova matter examined above.  
We assume here that the supernova matter is under the beta equilibrium without neutrinos to determine the net electron fraction as a function of density \cite{sch20}.  
We examine the dependence on the entropy per baryon of supernova matter to estimate the border to black hole formation in numerical simulations.  
The maximum mass for TM1m is smaller than that for TM1e with a difference of $\sim0.2M_{\odot}$.  
Therefore, one can expect an earlier timing of the black hole formation in the case of TM1m with respect to the case of TM1e for the same progenitor and accretion rate.  


%

\section{Numerical simulations
\label{sec:Sim}}

\subsection{Gravitational collapse of massive stars
\label{sec:massive_star}}

We perform numerical simulations of the gravitational collapse, core bounce, and post-bounce evolution of the central core of massive stars under spherical symmetry.  
We adopt the numerical code to solve the general relativistic neutrino-radiation hydrodynamics utilizing the solver of the Boltzmann equation for neutrinos \cite{sum05}.  
We perform a set of numerical simulations using the two EOS tables for TM1m and TM1e.  
It should be noted that we commonly use the EOS table of TM1e for the density region of non-uniform matter at densities below $10^{14}$~g\,cm$^{-3}$ so that we can extract the influence of uniform matter at high densities.  
We solve the neutrino distributions for 4 species ($\nu_e$, $\bar{\nu}_e$, $\nu_{\mu/\tau}$, $\bar{\nu}_{\mu/\tau}$) with 14 energy and 6 angle grids and matter distributions in the radial mass coordinate with 511 grids.  
The neutrino species $\nu_{\mu}$ and $\nu_{\tau}$ ($\bar{\nu}_{\mu}$ and $\bar{\nu}_{\tau}$) are collectively treated as $\nu_{\mu/\tau}$ ($\bar{\nu}_{\mu/\tau}$).  
The reaction rates via the weak interaction are implemented using the standard set with the extension for the nucleon-nucleon bremsstrahlung.  
We adopt the central Fe core of massive stars of 11.2M$_{\odot}$ by Woosley et al. in \cite{woo02} (WHW02), 15 and 40M$_{\odot}$ by Woosley and Weaver in \cite{woo95} (WW95) and 40M$_{\odot}$ by Sukhbold et al. in \cite{suk16} (S16).  
Except for the EOS, we adopt the same setting as used in the numerical simulations in \cite{sum19,sum22} for comparison with the results using the other EOS tables including the other sets of the Shen EOS.  

\subsubsection{Post bounce evolution
\label{sec:post_bounce}}

\begin{figure}
\includegraphics[width=5.5cm]{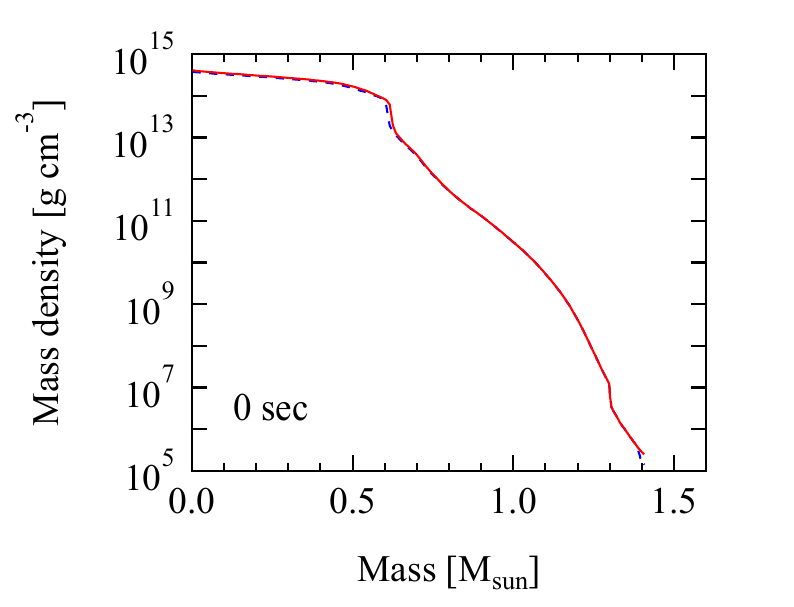}%
\includegraphics[width=5.5cm]{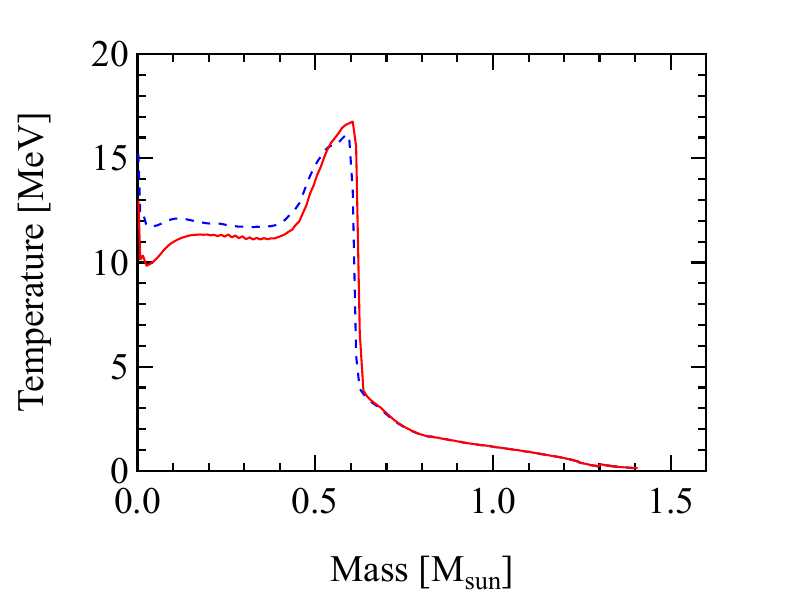}%
\includegraphics[width=5.5cm]{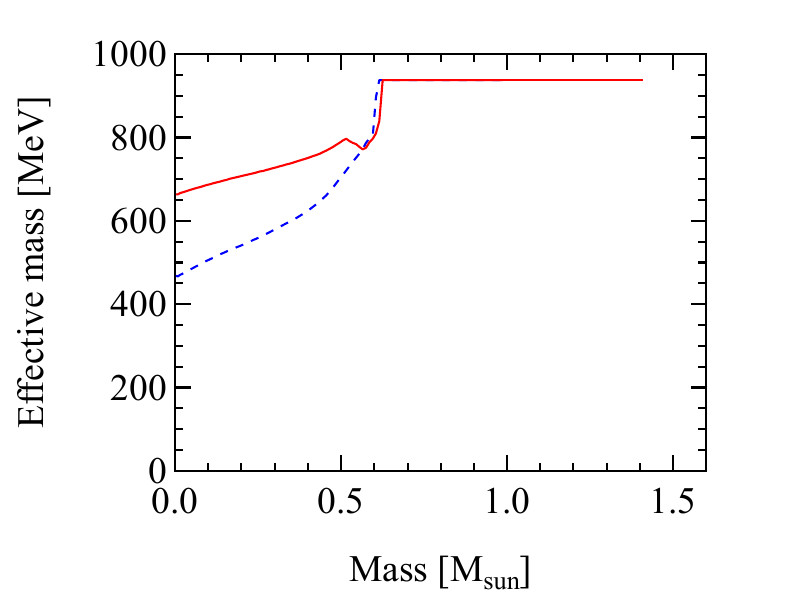}\\%
\includegraphics[width=5.5cm]{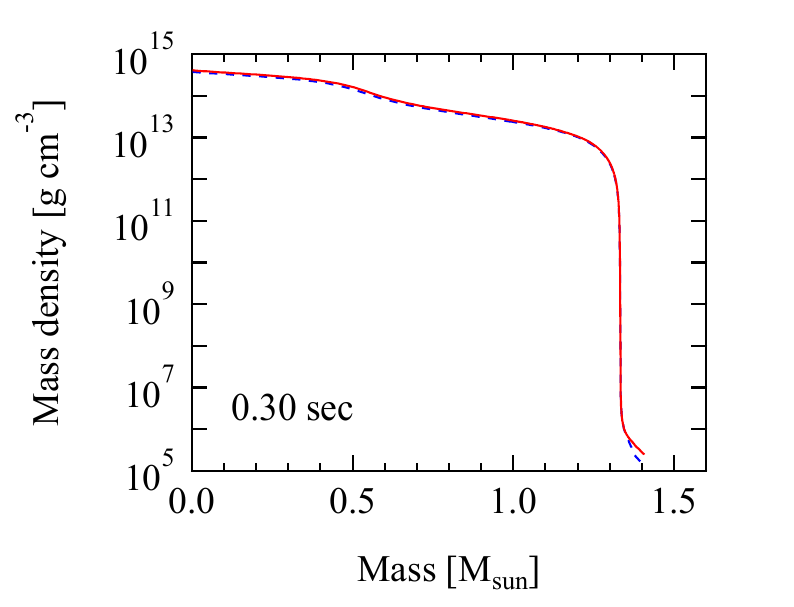}%
\includegraphics[width=5.5cm]{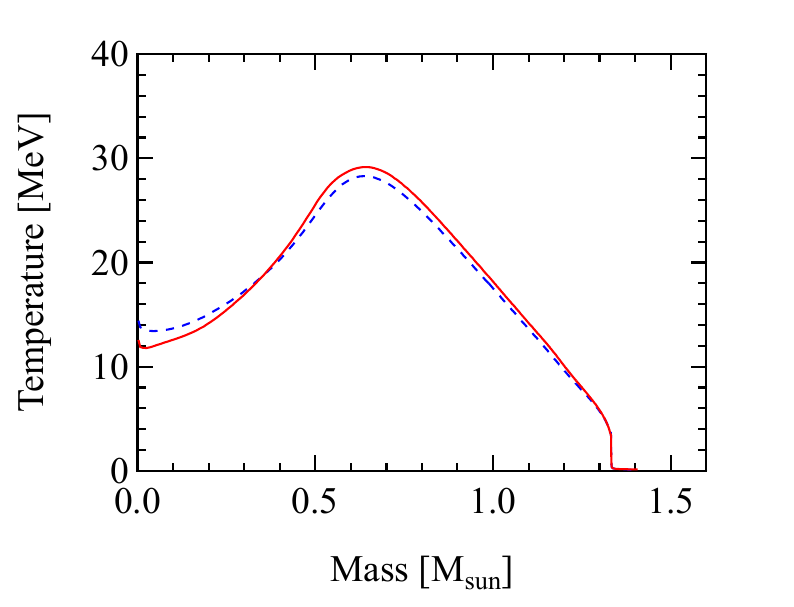}%
\includegraphics[width=5.5cm]{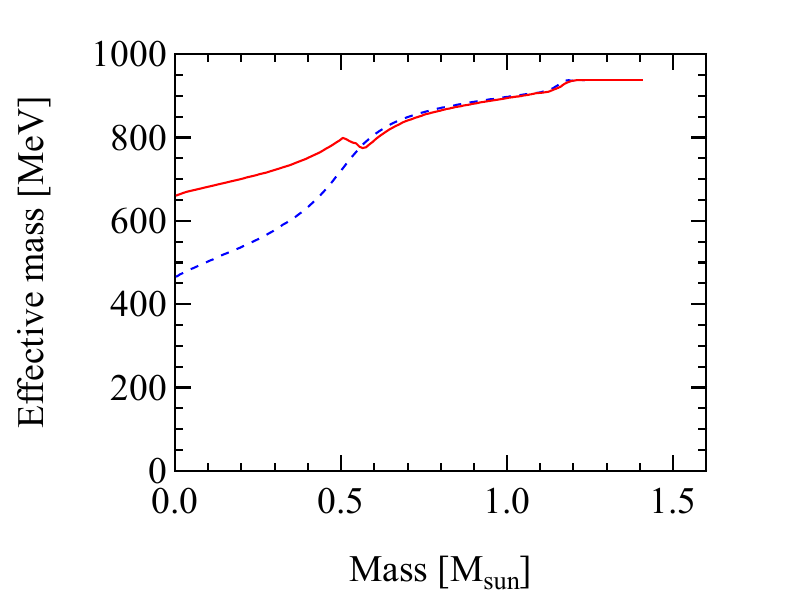}%
\caption{Profiles of the mass density, temperature, and effective mass are shown as functions of the radial mass coordinate in the left, middle, and right panels, respectively, in the case of TM1m (red, solid) and TM1e (blue, dashed).  
Upper and lower panels show the profiles at 0 and 0.30 s after the core bounce, respectively, from the massive star of 11.2M$_{\odot}$ by WHW02.  
\label{fig:WHW11M_profile}}
\end{figure}

In order to examine the influence of the effective mass on the properties of supernova core at and after the core bounce, we follow the time evolution of the collapse and bounce of massive stars of 11.2M$_{\odot}$ by WHW02 and 15M$_{\odot}$ by WW95 until 0.30 s after the core bounce (0 s).  
We show in Fig. \ref{fig:WHW11M_profile} the snapshots of the central core for the model of 11.2M$_{\odot}$ at 0 and 0.30 s after the core bounce.  
The initial position of shock wave for both models is nearly the same each other with similar values of the electron fraction and entropy per baryon at the core bounce.  
This is mainly due to the common usage of the TM1e EOS for non-uniform matter at low densities during the gravitational collapse.  
The difference between TM1e and TM1m EOSs appears after the matter becomes uniform at high densities.  
The central density for TM1m is $4.1\times10^{14}$~g\,cm$^{-3}$, which is slightly higher than $3.8\times10^{14}$~g\,cm$^{-3}$ for TM1e, reflecting the softness of TM1m compared to TM1e.  
The temperature for TM1m in the central region is lower than that for TM1e.  
This is the effect of different effective masses in the central region as seen in the upper right panel in Fig. \ref{fig:WHW11M_profile}.  
Larger effective mass for TM1m leads to lower temperature having nearly the same profiles of the entropy per baryon for both models.  
These differences remain the same at 0.30 s after the core bounce.  
We note that similar characteristics of differences and similarities are found in the results for the case of 15M$_{\odot}$.  
The influence on post bounce evolution through the increase of effective mass is consistent with the results found in \cite{sch19b} but the size of the difference is smaller.  
The difference will be larger at late stages through the contraction of proto-neutron stars as we show later in \S \ref{sec:PNS_cooling}.  


%

%
%

\subsubsection{Black hole formation
\label{sec:BH_formation}}

In order to examine the influence of the effective mass on the dynamics and neutrino signals for non-explosion cases, we follow the time evolution of the collapse and bounce of massive stars of 40M$_{\odot}$ up to the dynamical collapse to the black hole formation.  
After the core bounce, the shock wave stalls and recedes as a result of intense mass accretion.  
The proto-neutron star born at the center becomes increasingly massive, approaches the critical configuration with the maximum mass, and recollapses to the black hole.  
The density and temperature of the massive proto-neutron star become very high due to the configurations with large masses.  

\begin{figure}
\includegraphics[width=5.5cm]{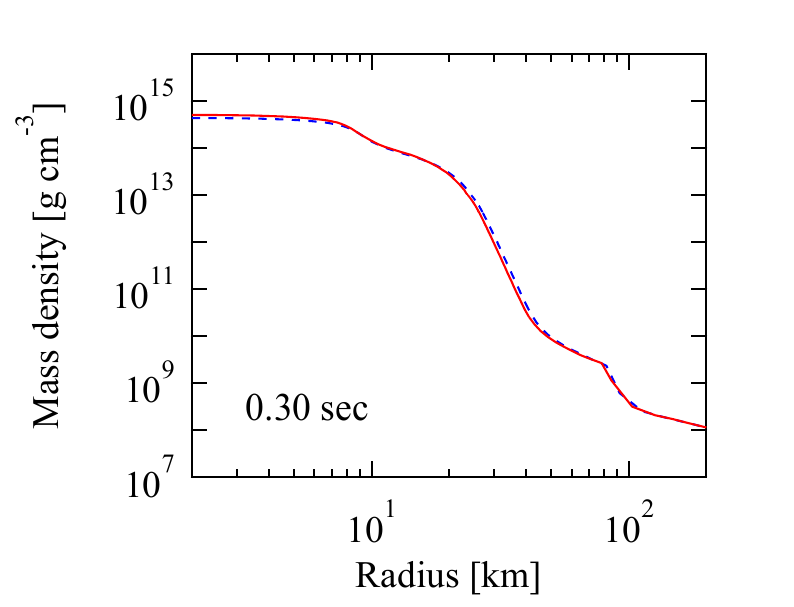}%
\includegraphics[width=5.5cm]{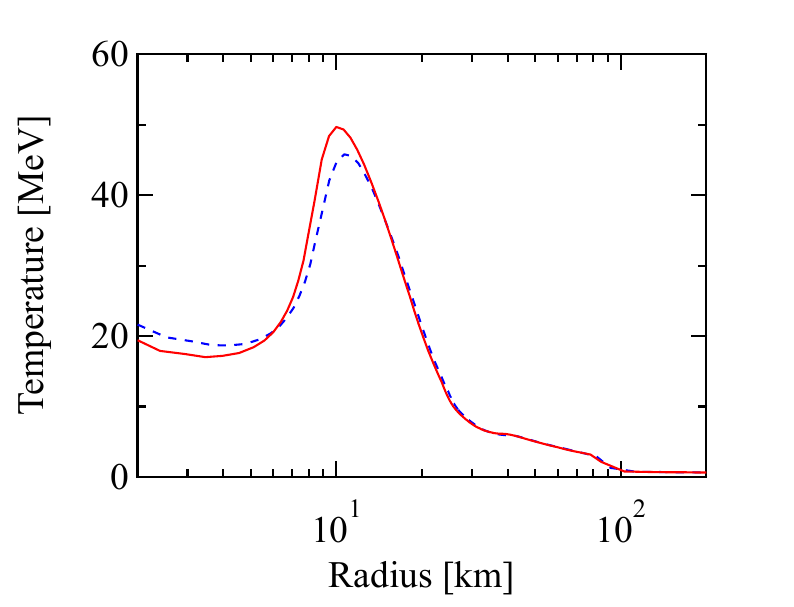}%
\includegraphics[width=5.5cm]{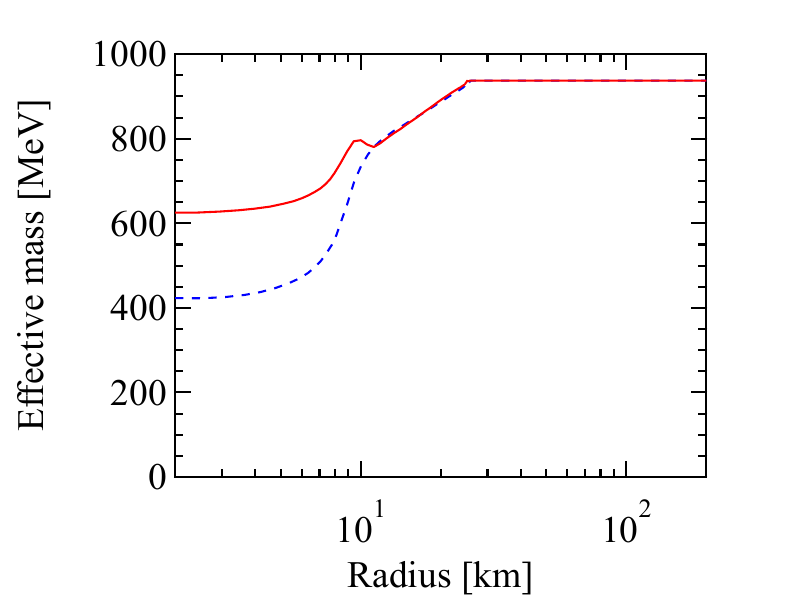}\\%
\includegraphics[width=5.5cm]{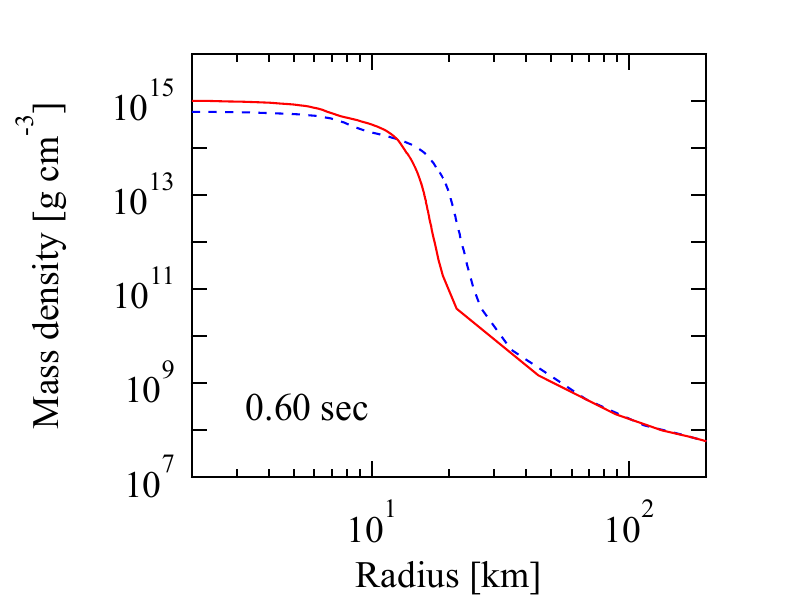}%
\includegraphics[width=5.5cm]{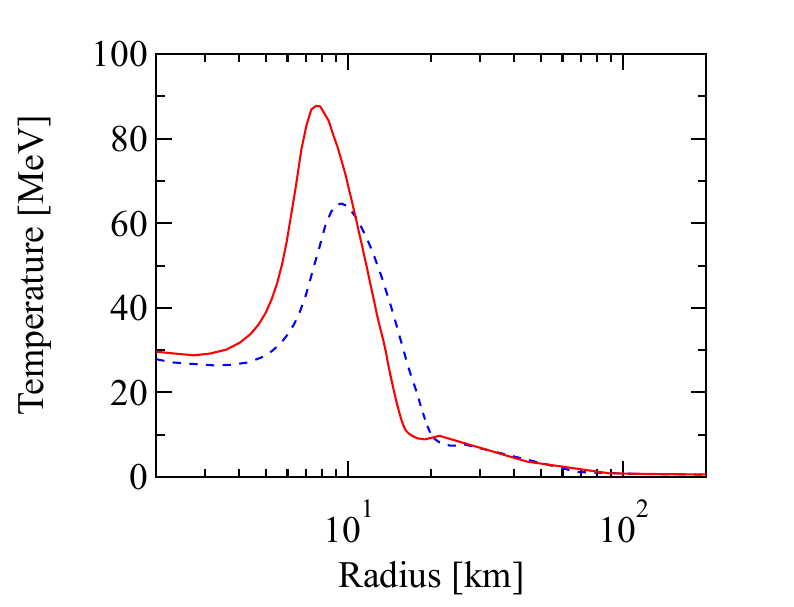}%
\includegraphics[width=5.5cm]{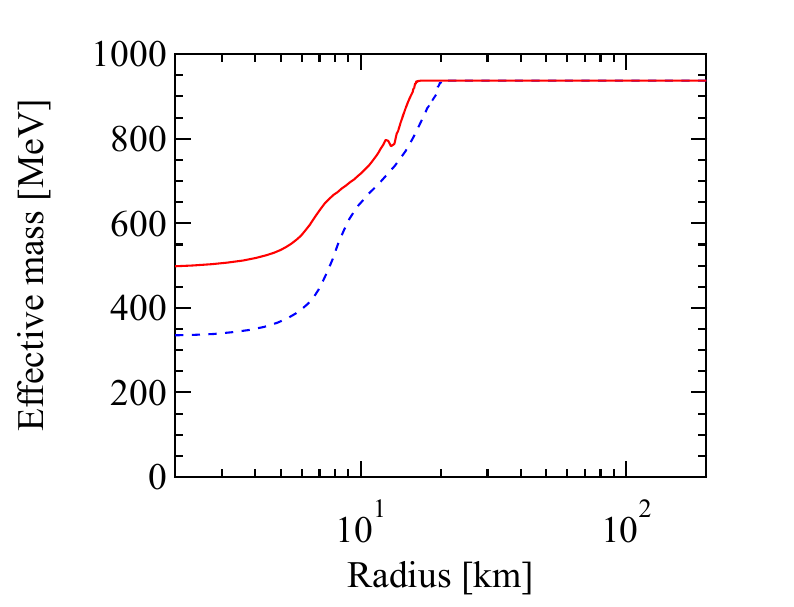}%
\caption{Profiles of supernova cores at 0.30 and 0.60 s after the core bounce from the massive star of 40M$_{\odot}$ by S16 
are shown in the upper and lower panels, respectively, in the case of TM1m (red, solid) and TM1e (blue, dashed).  
The mass density, temperature, and effective mass are shown as functions of the radius in the left, middle, and right panels, respectively.  
\label{fig:40M_profile}}
\end{figure}

We show in Fig. \ref{fig:40M_profile} profiles of the massive proto-neutron stars at 0.30 and 0.60 s after the core bounce from the massive star of 40M$_{\odot}$ by S16 
for the two models with TM1m and TM1e.  
At 0.30 s, the central density for TM1m is slightly higher and the central temperature for TM1m is lower than for TM1e.  
The temperature is lower when the effective mass is larger for the same value of entropy per baryon and density as seen before.  
The profiles are more different at 0.60 s since the mass of the proto-neutron star becomes more massive.  
The central density for TM1m is $1.0\times10^{15}$~g\,cm$^{-3}$, which is higher than $6.0\times10^{14}$~g\,cm$^{-3}$ for TM1e.  
The proto-neutron star for TM1m is more compact than for TM1e due to the softness of TM1m.  
Note that the baryon mass of the proto-neutron star is nearly the same value $\sim$2.34M$_{\odot}$.  
The temperature for TM1m is higher than for TM1e despite the larger effective mass of TM1m than TM1e with nearly the same profile of the entropy per baryon.  
This is because the compression of matter in the case of TM1m is more drastic than in the case of TM1e.  



\begin{figure}
\includegraphics[width=8cm]{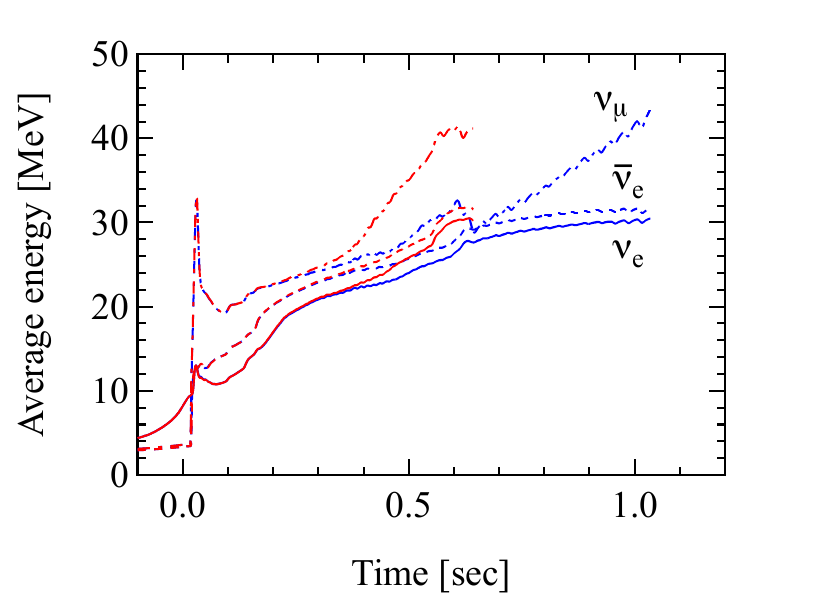}%
\includegraphics[width=8cm]{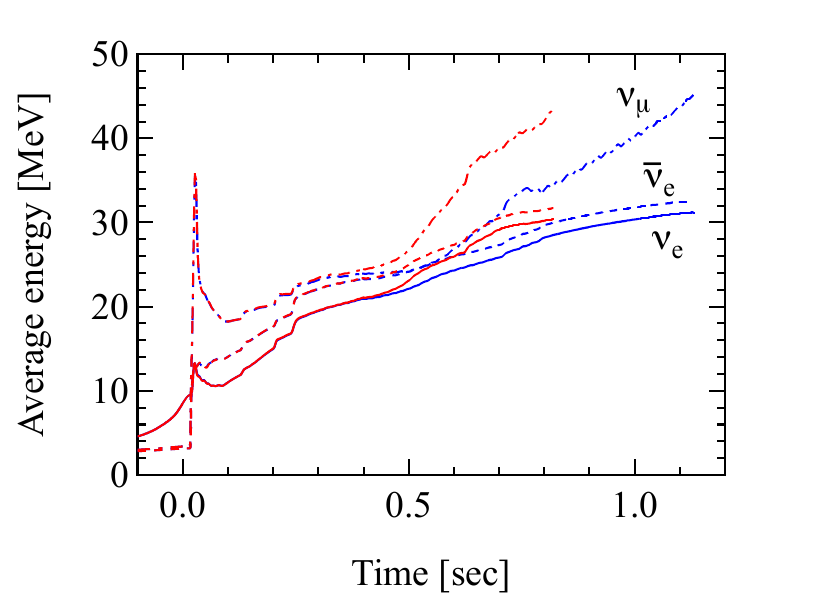}%
\caption{The average energies of neutrinos emitted from the collapse of 40M$_{\odot}$ by S16 
and 40M$_{\odot}$ by WW95 
are shown as functions of time after the bounce in the left and right panels, respectively.  
The three neutrino species, $\nu_e$, $\bar{\nu}_e$, and $\nu_{\mu}$, are shown by the solid, dashed, and dash-dotted lines, respectively, in the case of TM1m (red) and TM1e (blue).  
\label{fig:40M_Enu}}
\end{figure}

The emission of neutrinos from massive proto-neutron stars is influenced by the difference of the two EOS tables.  
We show in Fig. \ref{fig:40M_Enu} the time evolution of the average energy of neutrinos toward the black hole formation.  
The average energy of all species increases rapidly due to the increasing density and temperature in the massive proto-neutron star with the intense matter accretion.  
The neutrino emission is terminated at the black hole formation at different timings for the two models.  
Recollapse to the black hole occurs at 0.64 and 1.03 s after the core bounce in the case of 40M$_{\odot}$ by S16 
for TM1m and TM1e, respectively.  
The difference in the duration of neutrino emission is caused by the different maximum masses supported by the two sets of EOS, TM1m and TM1e, as seen in Fig. \ref{fig:protoNS_Mmax}.  

The behavior in the neutrino signal is similar also in the case of 40M$_{\odot}$ by WW95.  
The emission is terminated at 0.82 and 1.13 s for TM1m and TM1e, respectively, 
as shown in the right panel.  
The duration is different from the cases of 40M$_{\odot}$ by S16 
because the accretion rate is different depending on the density profiles of progenitors.  
Tendency having short duration with large effective mass is consistent with previous studies adopting EOS tables with Skyrme-type interaction \cite{sch20}.  

\subsection{Cooling of proto-neutron star
\label{sec:PNS_cooling}}

We perform numerical simulations of the thermal evolution of the proto-neutron star under spherical symmetry.  
We adopt the numerical code to solve general relativistic quasi-hydrostatic structure and neutrino transport by using the flux-limited diffusion approximation \cite{sum95c,suz94}.  
We construct the initial model based on the snapshots obtained from the dynamical simulations of core-collapse in the same way as in Refs. \cite{nak18,sum19,sum22}.  
It is to be noted that the profiles are originally obtained from the numerical simulation of the 15M$_{\odot}$ star by WW95 
with the Shen EOS using TM1.  
We adopt the distributions of entropy per baryon and electron fraction  as functions of the baryon mass coordinate in the central object with the baryon mass of 1.47M$_{\odot}$ at 0.30 s after the core bounce.  
We fix them to obtain the steady flow of neutrinos under the hydrostatic structure in the process of initial model constructions for each model.  
We perform a set of numerical simulations using the two EOS tables for TM1m and TM1e for high densities with common usage of TM1e for low densities.  
We solve the energy density and flux of neutrinos for three species ($\nu_e$, $\bar{\nu}_e$, $\nu_{\mu/\tau}$) with 25 energy grids and matter distributions with 99 radial grids.  
The neutrino species $\nu_{\mu}$, $\nu_{\tau}$, $\bar{\nu}_{\mu}$, and $\bar{\nu}_{\tau}$ are collectively treated as $\nu_{\mu/\tau}$.  
The reaction rates via the weak interaction corresponding to the one in the core-collapse simulations are implemented based on Bruenn’s rate and its extension.  
The same setting is used as in the numerical simulations in \cite{sum19,sum22} for comparison with previous results.  

\begin{figure}
\includegraphics[width=5.5cm]{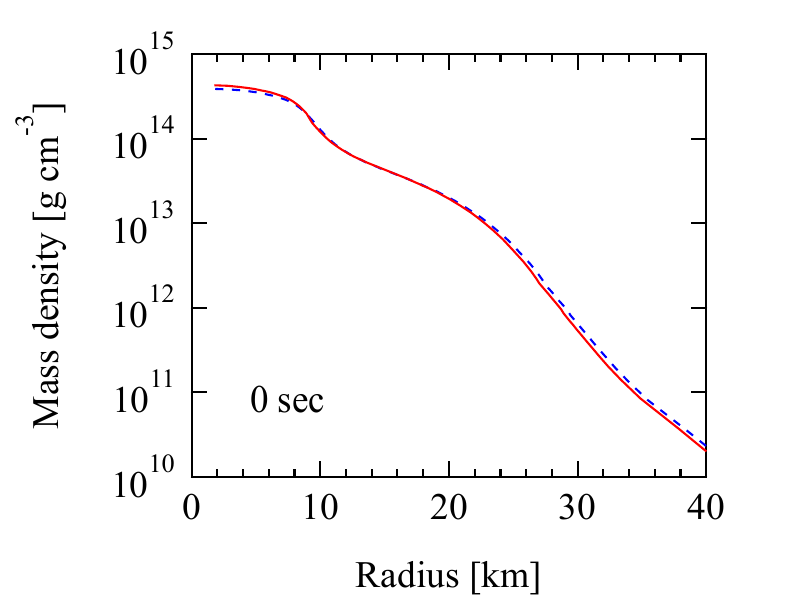}%
\includegraphics[width=5.5cm]{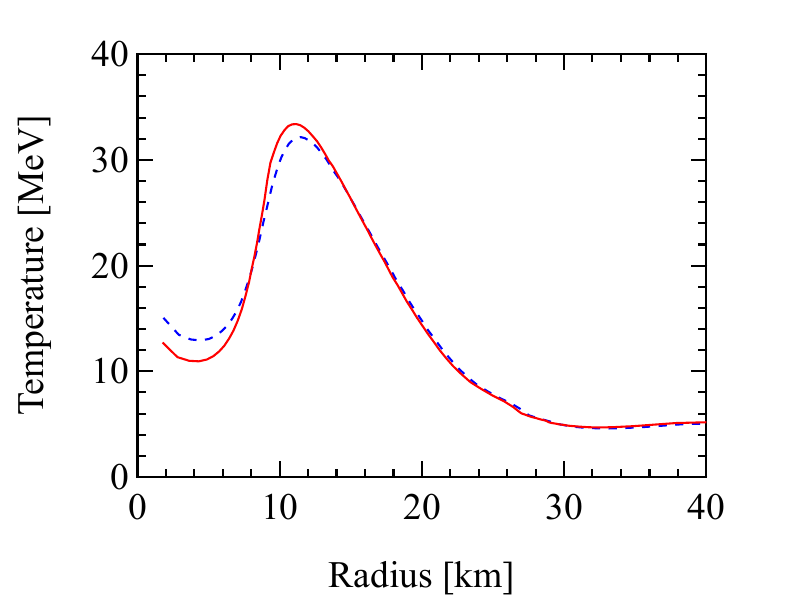}%
\includegraphics[width=5.5cm]{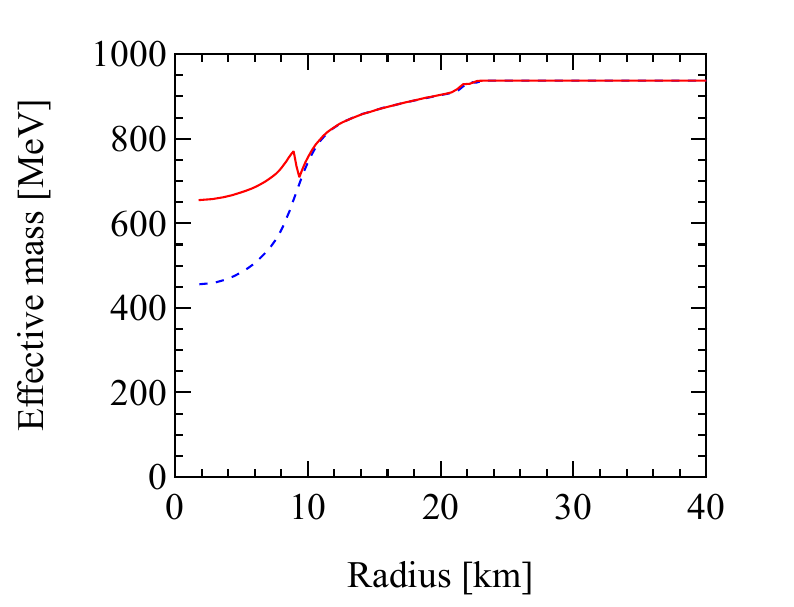}\\%
\includegraphics[width=5.5cm]{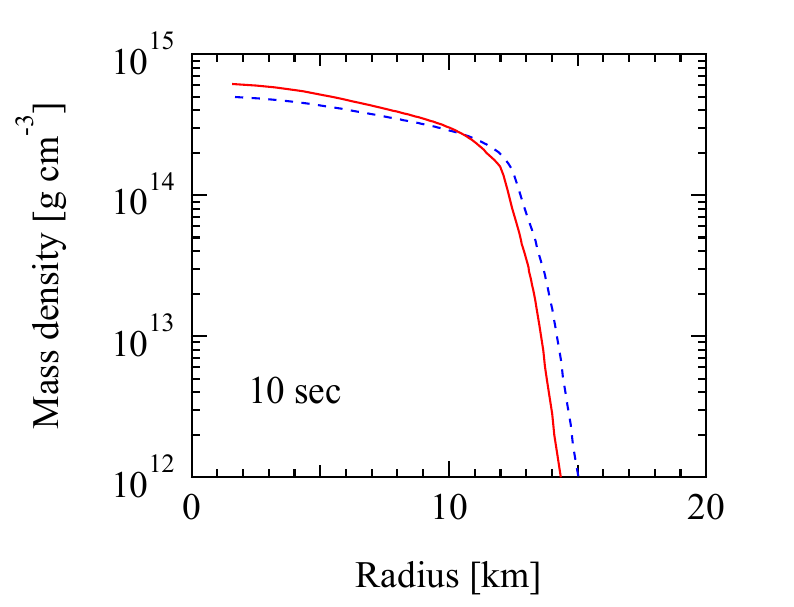}%
\includegraphics[width=5.5cm]{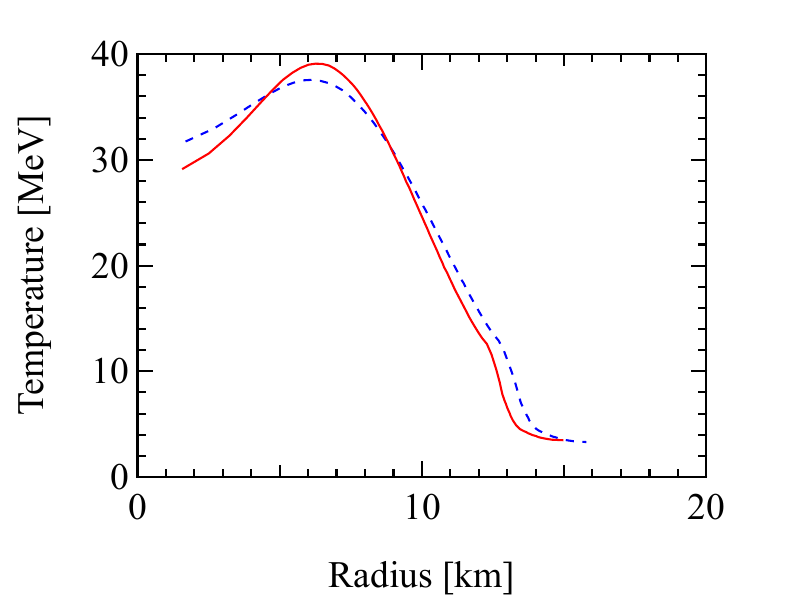}%
\includegraphics[width=5.5cm]{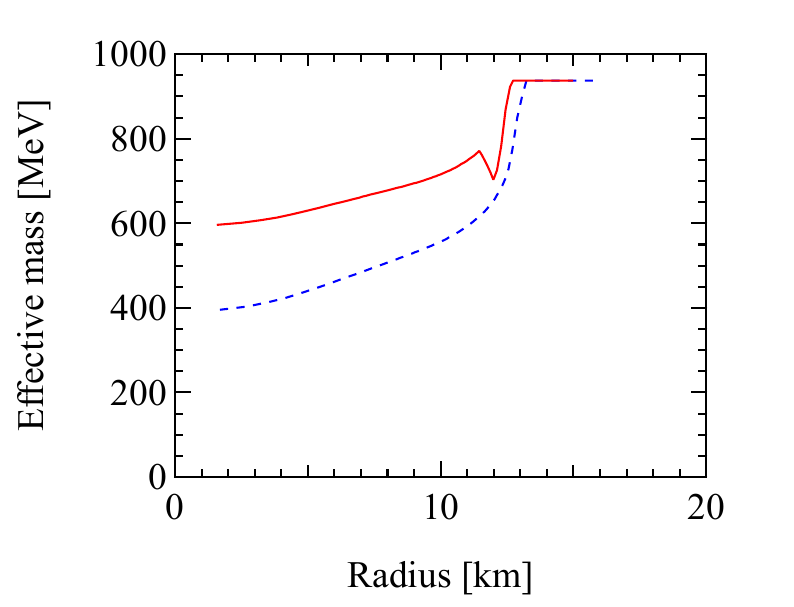}\\%
\includegraphics[width=5.5cm]{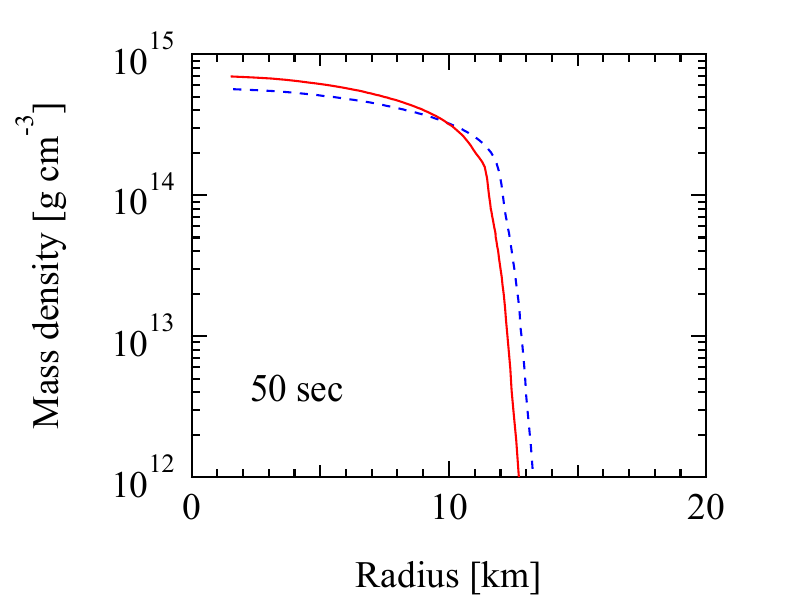}%
\includegraphics[width=5.5cm]{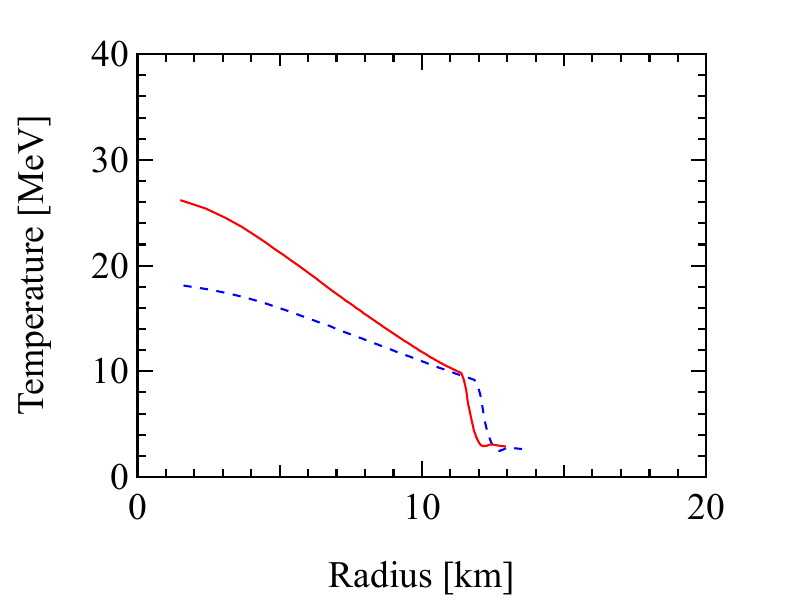}%
\includegraphics[width=5.5cm]{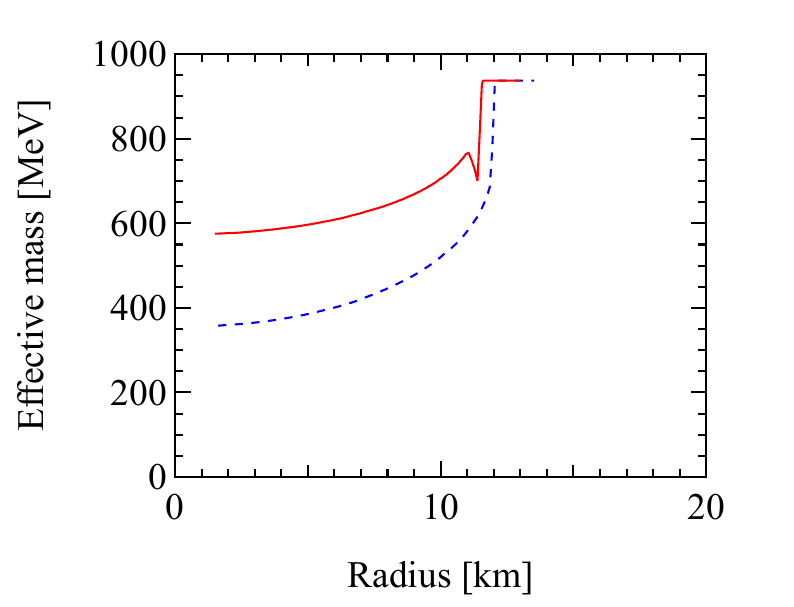}\\%
\caption{Profiles of the proto-neutron stars at 0, 10, and 50 s in the numerical simulations of thermal evolution are shown in the top, middle, and bottom panels, respectively, in the case of TM1m (red, solid) and TM1e (blue, dashed). The mass density, temperature, and effective mass are shown as functions of the radius in the left, middle, and right panels, respectively.  
\label{fig:PNS_profile}}
\end{figure}

We show in Fig. \ref{fig:PNS_profile} the profiles of the proto-neutron star at 0, 10, and 50 s in the numerical simulations of two models, TM1m and TM1e.  
The density and temperature at the initial condition (0 s) for the two models are similar to each other with a slight difference at the central part.  
This is consistent with the results of the post bounce evolution found in Fig. \ref{fig:WHW11M_profile}.  
The density and temperature become high due to compression at 10 s and thermal energy flows outward through the neutrino diffusion once the temperature profile becomes monotonic.  
The difference becomes clear at 50 s after the thermal evolution with neutrino emission.  
The proto-neutron star for TM1m is more compact with smaller radius due to the softness of TM1m EOS than in the case of TM1e EOS.  
Neutrinos are trapped more strongly and the temperature 
becomes higher as a result of the compression to the high densities in the case of TM1m.  
These differences result in slower diffusion and higher energy of neutrinos, as we show below.  

\begin{figure}
\includegraphics[width=8cm]{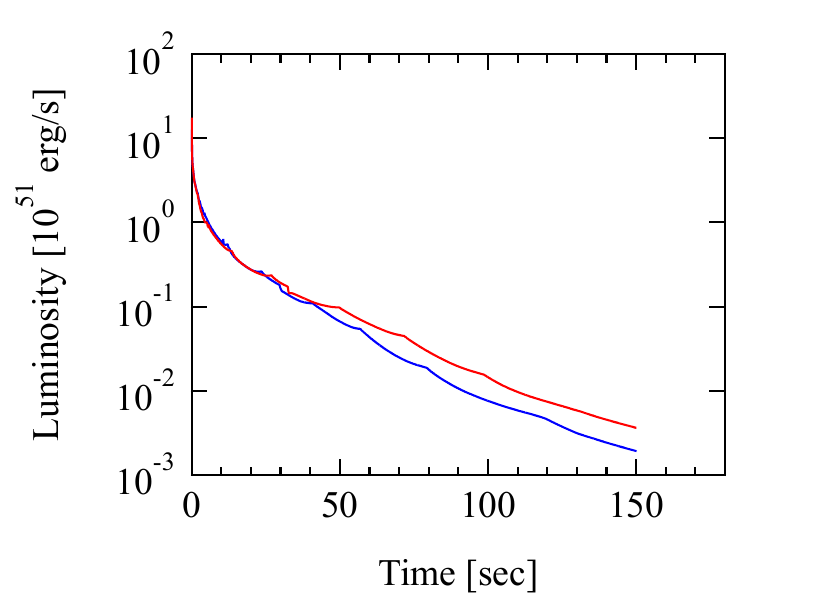}%
\includegraphics[width=8cm]{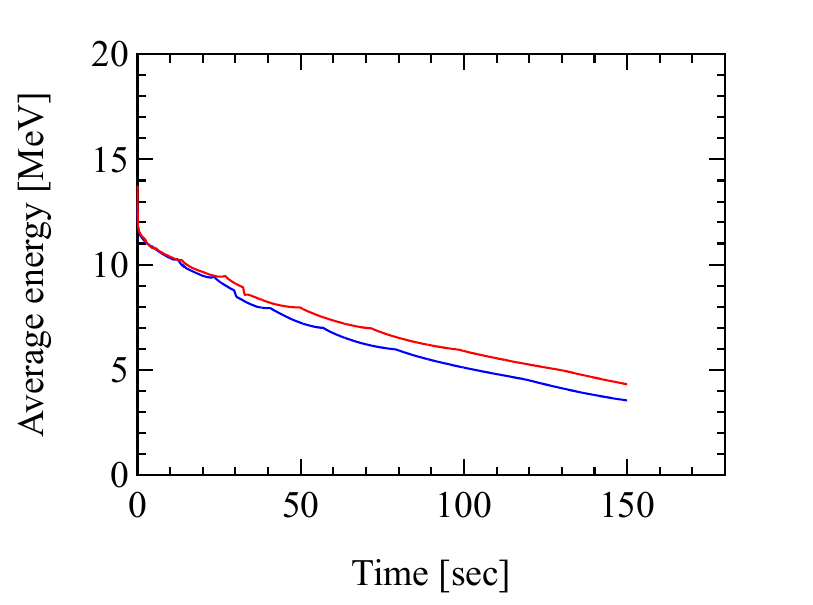}%
\caption{Time evolution of neutrino emission from the proto-neutron stars in the numerical simulations of thermal evolution is shown in the case of TM1m (red, solid) and TM1e (blue, dashed).  
Luminosity and average energy of $\bar{\nu}_e$ are shown as functions of time in the left and right panels, respectively.  
\label{fig:PNS_Nu}}
\end{figure}

We show in Fig. \ref{fig:PNS_Nu} the time evolution of neutrino emission from the proto-neutron stars in the two models TM1m and TM1e.  
The luminosity of neutrinos for TM1m remains higher than that for TM1e because of larger thermal energy and slow diffusion due to the high density.  
The average neutrino energies for TM1m are higher than that for TM1e due to the high temperature inside the proto-neutron star.  
The character of neutrino signals with high luminosity and energy reflects the difference in stiffness of EOS with different effective masses.  

The tendency of slow cooling evolution and neutrino emission due to the increase of effective mass is consistent with the one reported by \cite{nak19,nak20}.  
We note, however, that the value of effective mass is fixed as a constant and the thermal contribution of the EOS is separately evaluated by using the ideal Fermi gas in their study.  
In the current study, the thermodynamical quantities are consistently derived with the effective mass in the RMF theory.  
The thermal energy increases by the large effective mass through energy level density.  
In addition, the softness of EOS due to the large effective mass leads to high density and temperature.  
These facts enhance neutrino luminosities and the cooling time scale is prolonged as found in \cite{nak19,nak20}.  



\newpage

\section{Summary and Discussion
\label{sec:Summary}}

We investigated the influence of the effective mass of the hot and dense matter in core-collapse supernovae and formation of compact objects.  
We explored the effects of the effective mass in EOS derived by the relativistic many body framework, which has different properties from the non-relativistic approaches such as the Skyrme interaction used in previous studies.  
We utilized the two sets of the EOS table derived by the relativistic mean field (RMF) theory for comparison in numerical simulations.  
Our aim is to extract the effects of the effective mass by fixing other bulk properties at the saturation density by choosing the two interactions, TM1e and TM1m, of the relativistic mean field theory for the construction of the EOS tables.  

We compared the basic properties of the two EOS at zero temperature and examine differences in hot and dense matter.  
We demonstrated that the large effective mass leads to softening of the EOS through the balance between the attractive scalar and repulsive vector potentials.  
The behavior of the EOS is constrained by the effective lagrangian of the RMF theory and the effective mass is not a simple parameter but a determining factor of the attraction.  
Choosing a large value of the effective mass leads to the softening of the EOS due to the reduction of pressure through both the kinetic term and vector potentials and the resulting compact neutron stars.  
At high densities, the stiffening of the EOS is seen because of the growth of the repulsive vector potentials.  
This tendency is also seen at finite temperature for the hot and dense matter with leptons and radiations.  
The resulting maximum gravitational mass of the proto-neutron star at finite temperature is reduced for the large effective mass and alters the condition for the black hole formation.  
The large effective mass leads to an increase in the entropy per baryon through the energy level density.  
This results in a reduction of temperature in the adiabatic compression inside core-collapse supernovae and may affect the properties of trapped neutrinos.  

We performed numerical simulations on the post-bounce evolution of massive stars and the cooling of proto-neutron stars to examine the effects of different values of effective mass.  
High density and low temperature are observed with large effective mass in the central region, but their differences are not large up to 0.30 s after the bounce in the case of 11.2M$_{\odot}$ and 15M$_{\odot}$ stars.  
The difference becomes more drastic at later stage in the case of 40M$_{\odot}$ stars toward the black hole formation.  
The soft EOS with large effective mass leads to a compact proto-neutron star with high temperature and early recollapse to the black hole.  
The emission of neutrinos associated with the massive proto-neutron star is more energetic and terminated at early timing.  
The duration of the neutrino burst toward black hole formation is short with the large effective mass in the case of 40M$_{\odot}$ stars.  
The cooling of proto-neutron star is affected by the effective masses due to the softness during long evolution.  
The soft EOS with large effective mass leads to more compact proto-neutron stars with high density and temperature.  
This leads to a high luminosity and average energy of neutrinos emitted from the thermal energies stored inside.  
The duration of neutrino emission becomes long as a result of the slow time scale of diffusion of neutrinos at high densities.  


We comment on remaining issues on detailed processes in the current study.  
We adopted the common EOS tables for non-uniform matter at low density.  
Changes of the effective mass at low densities are smaller in general, but possible effects from the formation of nuclei during the collapse and in neutrino transport around the neutrinosphere can appear.  
The construction of the EOS table with the full coverage of density range is awaited and will be further applied to numerical simulations in the future.  
The effective mass may affect the neutrino reaction rates in the medium.  
Although the dependence on the effective mass is partly implemented in the current simulations, further studies are necessary in a consistent manner.  
Since the neutrino emission from supernovae is affected by the effective mass, observational neutrino signals can be used to probe their values in principle.  
It requires further studies on the event rates at the neutrino detectors by separating various effects of EOS other than the effective mass (see \cite{nak22} for example).  
It is also intriguing to explore the impact of the effective mass in the RMF frameworks on multi-dimensional dynamics since there is no explosion in most of numerical simulations under the spherical symmetry.  
\begin{acknowledgments}
This work is supported by 
Grant-in-Aid for Scientific Research
(24K00632, 24K07021, 24H02245, 25H01273, 26K07073, 26K07085) 
from the Ministry of Education, Culture, Sports, Science and Technology (MEXT), Japan.  
For providing high performance computing resources, 
Computing Research Center, KEK, 
JLDG on SINET of NII, 
Research Center for Nuclear Physics, Osaka University, 
Yukawa Institute of Theoretical Physics, Kyoto University, 
and 
Information Technology Center, University of Tokyo are acknowledged. 
Numerical studies in this work are supported by 
MEXT through 
"Program for Promoting Researches on the Supercomputer Fugaku" 
and 
the HPCI System Research Project.  
\end{acknowledgments}

\bibliography{sumi.bib}

\end{document}